\documentclass[epj]{svjour}
\usepackage{graphics}
\usepackage{bbm}
\usepackage{amsmath}

\def\empile#1\over#2{\mathrel{\mathop{\kern 0pt#1}\limits_{#2}}}
\newcommand{\slvarepsilon}{\raise.15ex\hbox{$/$}\kern-.53em\hbox{$\varepsilon$}}
\newcommand{\slL}{\raise.15ex\hbox{$/$}\kern-.53em\hbox{$L$}}
\newcommand{\slP}{\raise.15ex\hbox{$/$}\kern-.53em\hbox{$P$}}
\newcommand{\slp}{\raise.1ex\hbox{$/$}\kern-.63em\hbox{$p$}}
\newcommand{\slq}{\raise.1ex\hbox{$/$}\kern-.53em\hbox{$q$}}
\newcommand{\slv}{\raise.1ex\hbox{$/$}\kern-.63em\hbox{$v$}}
\newcommand{\slR}{\raise.15ex\hbox{$/$}\kern-.53em\hbox{$R$}}
\newcommand{\slQ}{\raise.15ex\hbox{$/$}\kern-.53em\hbox{$Q$}}
\newcommand{\slK}{\raise.15ex\hbox{$/$}\kern-.53em\hbox{$K$}}
\newcommand{\slk}{\raise.15ex\hbox{$/$}\kern-.53em\hbox{$k$}}
\newcommand{\slSigma}{\raise.15ex\hbox{$/$}\kern-.53em\hbox{$\Sigma$}}
\newcommand{\slcalP}{\raise.15ex\hbox{$/$}\kern-.63em\hbox{$\cal P$}}
\newcommand{\slA}{\raise.15ex\hbox{$/$}\kern-.73em\hbox{$A$}}
\newcommand{\slbfA}{\raise.15ex\hbox{$/$}\kern-.73em\hbox{${\imb A}$}}
\newcommand{\slpartial}{\raise.15ex\hbox{$/$}\kern-.53em\hbox{$\partial$}}
\newcommand{\sla}{\raise.15ex\hbox{$/$}\kern-.53em\hbox{$a$}}
\newcommand{\slb}{\raise.15ex\hbox{$/$}\kern-.53em\hbox{$b$}}
\newcommand{\slc}{\raise.15ex\hbox{$/$}\kern-.53em\hbox{$c$}}
\newcommand{\slC}{\raise.15ex\hbox{$/$}\kern-.63em\hbox{$C$}}

\def\p{{\boldsymbol p}}
\def\q{{\boldsymbol q}}
\def\l{{\boldsymbol l}}
\def\k{{\boldsymbol k}}

\def\x{{\boldsymbol x}}
\def\v{{\boldsymbol v}}
\def\u{{\boldsymbol u}}
\def\y{{\boldsymbol y}}

\def\z{{\boldsymbol z}}

\def\wt{\widetilde}

\begin{document}
\title{Quark production in high energy proton-nucleus collisions}
\author{H. Fujii\inst{1} 
\and F. Gelis\inst{2}
\and R. Venugopalan\inst{3}
}                     
\institute{Institute of Physics, University of Tokyo, Komaba, Tokyo 153-8902, Japan
\and CEA/DSM/SPhT, 91191 Gif-sur-Yvette, France 
\and Brookhaven National Laboratory, Nuclear Theory, Upton 11973, NY, USA
}
\date{Received: date / Revised version: date}
%
\abstract{In this note, we discuss the problem of quark-antiquark pair
  production in the framework of the color glass condensate. The
  cross-section can be calculated in closed form for the case of
  proton-nucleus collisions, where the proton can be considered to be
  a dilute object. We find that $k_\perp$-factorization is broken by
  rescattering effects.
} 
\maketitle
\section{Introduction}
The Color Glass Condensate (CGC) \cite{IancuV1,IancuLM3,Muell4}
provides a fra\-me\-work for studying hadronic collisions at high energy.
Following the ideas developed in the McLerran-Ve\-nu\-go\-pa\-lan Model
\cite{McLerV1,McLerV2,McLerV3,McLerV4}, it divides the partonic
degrees of freedom into hard color sources $\rho$ - partons with a
large momentum fraction $x$ that travel at almost the speed of light
and that can be considered as frozen over the typical interaction
times - and classical color fields $A^\mu$.  The latter represent the
small $x$ modes, which have a large occupation number and thus can be
described classically. The two types of modes are related by the fact
that the classical fields obey the Yang-Mills equations, with a source
term given by the hard color sources.

The hard sources are random variables (the partons at large $x$ come
in a different configuration in each collision), whose statistical
distribution is described by a functional density $W_{x_0}[\rho]$. The
subscript $x_0$ in this functional is the separation between the
degrees of freedom that are described as sources and those that are
described as fields. Upon changing this boundary, the functional
$W_{x_0}$ must change according to a renormalization group equation,
known as the JIMWLK equation
\cite{JalilKLW1,JalilKLW2,JalilKLW3,JalilKLW4,KovneM1,KovneMW3,Balit1,Kovch1,Kovch3,JalilKMW1,IancuLM1,IancuLM2}.
The initial condition for this evolution equation is in principle
non-perturbative. However, in the case of a large projectile like a
nucleus, it has been argued by McLerran and Venugopalan that a good
model for this initial condition is given by a Gaussian
\cite{McLerV1,McLerV2,McLerV3,McLerV4}. Such a Gaussian should remain
a reasonable description of the hard sources in a nucleus as long as
$x_0$ does not become too small.

In a collision between two hadronic projectiles, each projectile is
described by its own color source $\rho_{1,2}$, which correspond to
partons moving at the speed of light in opposite directions. The first
step in studying this collision process is to solve the Yang-Mills
equations in the presence of these two source terms. So far, this has
been achieved analytically only in the case where at least one of the
sources is assumed to be a weak source and treated perturbatively
(i.e. the solution of Yang-Mills equations is expanded order by order
in powers of this weak source) \cite{DumitM1,BlaizGV1}. This situation
of course prevails in the collision between two small, non saturated,
projectiles, like two protons at not too small $x$. It is also
relevant in the case of a small and a large projectile, like
proton-nucleus collisions, or deuteron-nucleus collisions as performed
at RHIC. The solution of Yang-Mills equations in this asymmetrical
situation will be reviewed in section \ref{sec:fields}. The case of
nucleus-nucleus collisions, where none of the color sources can be
considered to be weak, is at the moment out of reach of analytic
calculations, but can be solved numerically
\cite{KrasnV1,KrasnV2,KrasnNV1,KrasnNV2,Lappi1}.

Once the classical gauge fields are known, physical cross-sections can
be calculated. It is straightforward to obtain the gluon
multiplicity\footnote{In the case where both sources are strong, and
  thus the classical gauge fields are also strong fields, it is much
  simpler to evaluate inclusive quantities like multiplicities instead
  of cross-sections for more exclusive processes. Indeed, the latter
  involves the calculation of time-ordered amplitudes in the presence
  of the classical gauge field. The fact that this gauge field is time
  dependent in a collision process renders this problem very difficult
  in general due to the presence of vacuum-to-vacuum diagrams
  \cite{BaltzGMP1}. On the contrary, the calculation of multiplicities
  only involves retarded amplitudes in the classical field, for which
  there are no vacuum-to-vacuum diagrams.} at the classical level,
since it is simply a matter of doing a Fourier decomposition of the
gauge fields.  In order to obtain the multiplicity of produced
quark-antiquark pairs, one needs the retarded propagator of a quark in
the presence of the previously determined classical gauge field.
Again, in situations where the background gauge field is only known
numerically, the quark propagator can only be computed 
numerically as well \cite{Dietr1,Dietr2,GelisKL1}. For collisions
between a small and a large projectile however, one can obtain the
quark production cross-section analytically, as was already the case
for the classical color fields.  The result of this calculation will
be presented in section \ref{sec:cs}.

Finally, as a last step involved in calculating an observable relevant
for hadronic collisions, one must perform the average over the
configurations of the hard color sources, weighted by a factor
$W[\rho]$ for each projectile. This average cannot be performed
analytically in general, except in some special cases like when
$W[\rho]$ is a Gaussian functional. The source averages that we need
in the expression of the quark production cross-section will be given
in section \ref{sec:avg}. 

At this point, the quark production cross-section is expressed as a
multi-dimensional integral that cannot be further simplified
analytically. One observes that $k_\perp$-fac\-to\-ri\-za\-tion is broken
by re-scattering effects, contrary to what happened in the simpler case
of gluon production. Some implications of this formula are discussed
for single quark production and for quarkonium production in section
\ref{sec:discussion}.

\section{Gauge fields}
\label{sec:fields}
When studying the collision of two hadrons in the framework of the
Color Glass Condensate, the first step is to determine the classical
color field created in this collision\footnote{An exception to this
  is the case where one treats one of the two hadrons  using the
  standard collinear factorization instead of a description based on
  the CGC \cite{DumitJ1,DumitJ2,GelisJ1,GelisJ2,GelisJ3,Jalil1}. In this
  case, one needs only the gauge field produced by the other hadron
  alone (which is of course much simpler).}. One has to solve the
Yang-Mills equations,
\begin{equation}
\left[D_\mu,F^{\mu\nu}\right]=J^\nu\; .
\end{equation}
For these equations to be compatible with the constraints provided by
the Jacobi identity, the color current $J^\nu$ that appears in the
r.h.s. must be covariantly conserved:
\begin{equation}
\left[D_\nu,J^\nu\right]=0\; .
\end{equation}
This condition implies that the current in general receives
corrections to all orders in the color sources. At lowest order in the
sources, it is given by
\begin{equation}
J^\nu(x)=
g\delta^{\nu +}\delta(x^-)\rho_{1}(\x_\perp)
+
g\delta^{\nu -}\delta(x^+)\rho_{2}(\x_\perp)\; ,
\end{equation}
where $\rho_1$ and $\rho_2$ are the distributions of color charge in
the hadron moving to the right and to the left respectively. The
previous three equations must be supplemented by a gauge condition. In
our work, we have used the Lo\-renz-co\-va\-riant gauge
\begin{equation}
\partial_\mu A^\mu =0\; .
\end{equation}

Thus far, the solution of this set of equations is not known to all
orders in both $\rho_1$ and $\rho_2$ (although numerical solutions
have been obtained). The best one can achieve is to obtain the
solution to first order in $\rho_1$ (the source describing the proton)
and all orders in $\rho_2$ (the source describing the nucleus). In the
Lorenz gauge, this solution has a nice and compact form, which reads, in Fourier
space \cite{BlaizGV1},
\begin{eqnarray}
 A^\mu(q)&=& A_p^\mu(q)
+\frac{ig}{q^2+iq^+\epsilon}
\nonumber\\
&\times&
\int\frac{d^2\k_{1\perp}}{(2\pi)^2}
\bigg\{
C_{_{U}}^\mu(q,\k_{1\perp})\, 
\left[U(\k_{2\perp})-(2\pi)^2\delta(\k_{2\perp})\right]
\nonumber\\
&&
+
C_{_{V}}^\mu(q)\, 
\left[V(\k_{2\perp})-(2\pi)^2\delta(\k_{2\perp})\right]
\bigg\}\frac{{\rho_1}(\k_{1\perp})}{k_{1\perp}^2}\; .
\label{eq:field-1}
\end{eqnarray}
In this formula, the first term is the color field of the proton
alone:
\begin{equation}
A_p^\mu(q)=2\pi g
\delta^{\mu+}\delta(q^-)\frac{\rho_1(\q_\perp)}{q_\perp^2}\; .
\end{equation}
In the second term, $\k_{1\perp}$ is the momentum coming from the
proton and $\k_{2\perp}$, defined as $\k_{2\perp}\equiv
\q_\perp-\k_{1\perp}$, is the momentum coming from the nucleus. The
4-vectors\footnote{They are related to Lipatov's effective vertex
$C_{_{L}}^\mu$ by $C_{_{U}}^\mu+\frac{1}{2}C_{_{V}}^\mu=C_{_{L}}^\mu$.}
$C_{_{U}}^\mu$ and $C_{_{V}}^\mu$ have been introduced in \cite{BlaizGV1},
and $U$, $V$ are the Fourier transforms of Wilson lines in the
adjoint representation of $SU(N)$:
\begin{eqnarray}
&&
U(\x_\perp)\equiv {\cal P}_+ \exp\left[ig\int_{-\infty}^{+\infty}
dz^+ A_{_A}^-(z^+,\x_\perp)\cdot T
\right]\; ,\nonumber\\
&&V(\x_\perp)\equiv {\cal P}_+ \exp\left[i\frac{g}{2}\int_{-\infty}^{+\infty}
dz^+ A_{_A}^-(z^+,\x_\perp)\cdot T
\right]\; ,
\end{eqnarray}
where $A_{_A}^-$ is the gauge field of the nucleus alone. The $T^a$
are the generators of the adjoint representation of $SU(N)$ and ${\cal
  P}_+$ denotes a ``time ordering'' along the $z^+$ axis. The
peculiarity of this result is that it contains a Wilson line ($V$)
with an unusual factor $1/2$ in the exponent. Such a Wilson line must
be an artifact of the gauge we use, and its cancellation from physical
quantities is a non-trivial test of the gauge invariance of the final
result\footnote{The gauge field produced in pA collisions has also
  been determined in the Fock-Schwinger gauge $x^+A^-+x^-A^+=0$
  \cite{DumitM1} and in the gauge $A^+=0$ \cite{Metha1}. It has been
  checked explicitly that although the gauge fields are not the same
  in these other gauges, they all lead to the very same expression for
  the multiplicity of produced gluons.}.

\section{Quark production}
\label{sec:cs}
\subsection{Pair production amplitude}
Once the gauge fields have been obtained, the probability of producing
a quark-antiquark pair in the collision is given by\footnote{As shown
in \cite{BaltzGMP1}, this formula is in general incomplete because it
does not include the contribution of vacuum-to-vacuum diagrams, which
is in principle required by unitarity. However, in our approximation
where only the leading order in the source $\rho_1$ is kept, this
correction can be safely neglected. Note also that in this
approximation, it would be equivalent to use the retarded quark
propagator, instead of the time-ordered propagator.}:
\begin{equation}
\omega_\q\frac{d P_{_{Q\overline{Q}}}}{d^3\q}
=
\frac{1}{16\pi^3}
\int\frac{d^3\p}{(2\pi)^32\omega_\p}
\left|
\overline{u}(\q)
T(q,-p)
v(\p)
\right|^2\; ,
\end{equation}
where $\omega_\q\equiv \sqrt{\q^2+m^2}$ is the on-shell energy of a
quark, and $T(q,-p)$ the time-ordered propagator (amputated of its
external legs) of a quark in the presence of the previously obtained
color field, with incoming momentum $-p$ and outgoing momentum
$q$. This probability must be integrated over the impact parameter of
the collision in order to obtain the pair production cross-section.

As explained in \cite{BlaizGV2}, there is one technical complication
when using the gauge field given in eq.~(\ref{eq:field-1}) in order to
calculate the pair production amplitude. Indeed, it turns out that one
of the terms in the gauge field is proportional to $\delta(x^+)$ in
coordinate space, which means that such a term allows the vertex where
the pair is being produced to be located inside the
nucleus. Practically, this means that this term must be considered
separately, because its contribution can only be calculated by
smearing out the thickness of the nucleus. When this is done properly,
one obtains a pair production amplitude in which the spurious Wilson
line $V$ does not appear any longer:
\begin{eqnarray}
&&\!\!\!\!\!\!\!\!\!\!\!\!
{\cal M}_{_{F}}(\q,\p)\!=\!g^2\!\int\!\frac{d^2\k_{1\perp}}{(2\pi)^2}
\frac{d^2\k_\perp}{(2\pi)^2}
\frac{\rho_{p,a}(\k_{1\perp})}{k_{1\perp}^2}
\!\int\!\! d^2\x_\perp d^2\y_\perp
\nonumber\\
&&\times
e^{i\k_\perp\cdot\x_\perp}
e^{i(\p_\perp\!+\!\q_\perp\!-\!\k_\perp\!-\!\k_{1\perp})\cdot\y_\perp}
\nonumber\\
&&\times
\overline{u}(\q)\Big\{ T_{q\bar{q}}(\k_{1\perp},\k_{\perp})
[{\wt U}(\x_\perp)t^a {\wt U}^\dagger(\y_\perp)]
\nonumber\\
&&\qquad\quad+T_{g}(\k_{1\perp})[t^bU^{ba}(\x_\perp)]\Big\} v(\p)
\; ,\nonumber\\
&&
\label{eq:Mf-final-1}
\end{eqnarray}
where we denote:
\begin{eqnarray}
&&T_{q\bar{q}}(\k_{1\perp},\k_{\perp})\equiv
\nonumber\\
&&\equiv
\frac{\gamma^+(\slq-\slk+m)\gamma^-(\slq-\slk-\slk_1+m)\gamma^+}
{2p^+[(\q_\perp\!-\!\k_\perp)^2+m^2]+2q^+[(\q_\perp\!-\!\k_\perp\!-\!\k_{1\perp})^2+m^2]}
\nonumber\\
&&T_{g}(\k_{1\perp})\equiv 
\frac{\slC_{_{L}}(p+q,\k_{1\perp})}{(p+q)^2}
\; .
\label{eq:Tqqbar-Tg}
\end{eqnarray}
In eq.~(\ref{eq:Mf-final-1}), ${\wt U}$ is the same Wilson line as
$U$, except that it is now in the fundamental representation of the
gauge group.

\subsection{Pair cross-section}
At this point, it is a simple matter of squaring this amplitude,
performing the average over the color sources in both projectiles, and
integrating over the impact parameter of the collision in order to
obtain the cross-section for pair production. This gives
\begin{eqnarray}
&&\frac{d\sigma_{_{Q\overline{Q}}}}{d^2\q_\perp dy_q d^2\p_\perp dy_p}=
\nonumber\\
&&=
\frac{\alpha_s^2 N}{8\pi^4 (N^2-1)}
\int\limits_{\k_{1\perp},\k_{2\perp}}
\!\!\!
\frac{\delta(\p_\perp\!+\!\q_\perp\!-\!\k_{1\perp}\!-\!\k_{2\perp})}
{\k_{1\perp}^2 \k_{2\perp}^2}
\nonumber\\
&&
\times\Big\{
\int_{\k_\perp,\k_\perp^\prime}\!
{\rm tr}
\Big[(\slq\!+\!m)T_{q\bar{q}}(\slp\!-\!m)
T_{q\bar{q}}^{*\prime}\Big]
\phi_{_A}^{q\bar{q},q\bar{q}}(\k_{2\perp};\k_\perp,\k_\perp^\prime)
\nonumber\\
&&
+\int_{\k_\perp}
\!
{\rm tr}
\Big[(\slq\!+\!m)T_{q\bar{q}}(\slp\!-\!m)
T_{g}^{*}\Big]
\phi_{_A}^{q\bar{q},g}
(\k_{2\perp};\k_\perp)+{\rm h.c.}
\nonumber\\
&&\;
+{\rm tr}
\Big[(\slq\!+\!m)T_{g}(\slp\!-\!m)T_{g}^{*}\Big]
\phi_{_A}^{g,g}(\k_{2\perp})
\Big\}
\varphi_p(\k_{1\perp})\; ,
\label{eq:cross-section-qqbar}
\end{eqnarray}
The function $\varphi_p$, and the various $\phi_{_{A}}$'s are
correlators of color sources, defined as follows \cite{BlaizGV2}:
\begin{eqnarray}
&&\varphi_p(\l_\perp)\equiv \pi^2 R_p^2 g^2 l_\perp^2 
\int_{\x_\perp}
e^{i\l_\perp\cdot\x_\perp}\left<\rho_{1,a}(0)\rho_{1,a}(\x_\perp)\right>\; ,
\nonumber\\
&&\phi_{_{A}}^{g,g}(\l_{\perp})\equiv \frac{\pi^2 R_{_A}^2 l_\perp^2}{g^2 N}
\int_{\x_\perp}
e^{i\l_\perp\cdot\x_\perp}
\left<U(0)U^\dagger(\x_\perp)\right>_{aa}\; ,
\nonumber\\
&&\phi_{_A}^{q\bar{q},g}(\l_\perp;\k_\perp)\equiv
\frac{2\pi^2 R_{_A}^2 l_\perp^2}{g^2 N}
\int_{\x_\perp,\y_\perp}\!\!\!\!\!\!\!\!\!\!
e^{i(\k_\perp\cdot\x_\perp+(\l_\perp-\k_\perp)\cdot\y_\perp)}
\nonumber\\
&&\qquad\qquad\qquad\times
{\rm tr}\left<
{\wt U}(\x_\perp)t^a {\wt U}^\dagger(\y_\perp) t^b U_{ba}(0)
\right>\; ,
\nonumber\\
&&\phi_{_A}^{q\bar{q},q\bar{q}}(\l_\perp;\k_\perp,\k_\perp^\prime)\equiv
\frac{2\pi l_\perp^2}{g^2 N}
\nonumber\\
&&\quad\times
\int_{\x_\perp,\y_\perp,\u_\perp,\v_\perp}
\!\!\!\!\!\!\!\!\!\!\!\!\!\!\!\!\!\!\!\!\!\!\!\!
e^{i(\k_\perp\cdot \x_\perp-\k_\perp^\prime\cdot \u_\perp)}
e^{i(\l_\perp-\k_\perp)\cdot \y_\perp}
e^{-i(\l_\perp-\k_\perp^\prime)\cdot\v_\perp}
\nonumber\\
&&\quad\times
{\rm tr}\left<
{\wt U}(\x_\perp)t^a {\wt U}^\dagger(\y_\perp){\wt U}(\v_\perp)t^a
{\wt U}^\dagger(\u_\perp)
\right>\; .
\end{eqnarray}
The momentum assignments for the various distributions we have defined
are illustrated below:
\setbox1=\hbox to 3cm{\resizebox*{3cm}{!}{\includegraphics{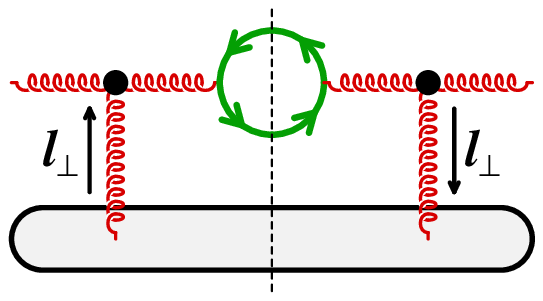}}}
\setbox2=\hbox to 3cm{\resizebox*{3cm}{!}{\includegraphics{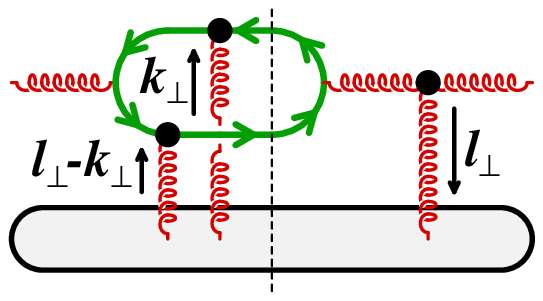}}}
\setbox3=\hbox to 3cm{\resizebox*{3cm}{!}{\includegraphics{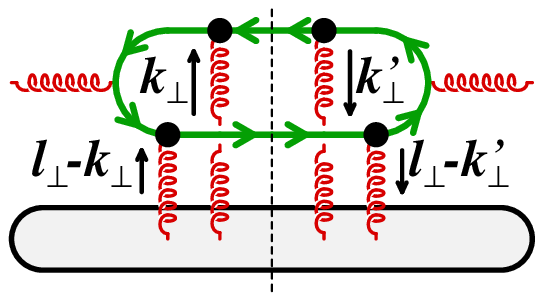}}}
\begin{eqnarray}
\phi_{_A}^{g,g}(\l_\perp)&\equiv&\qquad
\raise -5mm\box1\quad ,
\nonumber\\
\phi_{_A}^{q\bar{q},g}(\l_\perp;\k_\perp)&\equiv&\qquad
\raise -5mm\box2\quad ,
\nonumber\\
\phi_{_A}^{q\bar{q},q\bar{q}}(\l_\perp;\k_\perp,\k_\perp^\prime)&\equiv&\qquad
\raise -5mm\box3\quad .
\end{eqnarray}

It is not possible to write this cross-section in terms of a single
distribution describing the nucleus, and therefore a strict
$k_\perp$-factorization is impossible to achieve\footnote{Let us recall 
that in the case of gluon production, the cross-section can
  be expressed in terms of the function $\phi_{_A}^{g,g}$ only
  \cite{KovchT1,DumitM1,BlaizGV1}.  Therefore, even though this
  function is not the canonical unintegrated gluon distribution of the
  nucleus, one has a cross-section which is exactly
  $k_\perp$-factorizable.}.  However, a relaxed form of
$k_\perp$-factorization remains true, termed 
``non-linear $k_\perp$-factorization'' 
in \cite{NikolS1,NikolSZ1}, provided one
introduces three distinct functions describing the content of nucleus.
The three $\phi_{_A}$'s are not entirely independent, since they are
related by some sum rules \cite{BlaizGV2}, but they nevertheless probe
different correlations among the color sources contained in the
nucleus.

\subsection{Single quark cross-section}
It is also possible to integrate out the phase-space of the antiquark
in order to obtain the inclusive single quark cross-section~\footnote{Single quark production has also been studied by 
Tuchin for Gaussian correlators. Unfortunately, his results are in co-ordinate space and cannot be easily compared to~\cite{Tuchin}}. One
obtains \cite{BlaizGV2}
\begin{eqnarray}
&&\frac{d\sigma_{_Q}}{d^2\q_\perp dy_q}=
\frac{\alpha_s^2 N}{8\pi^4 (N^2-1)}\int\frac{dp^+}{p^+}
\int_{\k_{1\perp},\k_{2\perp}}
\frac{1}{\k_{1\perp}^2 \k_{2\perp}^2}
\nonumber\\
&&
\times\Big\{
{\rm tr}
\Big[(\slq\!+\!m)T_{q\bar{q}}(\slp\!-\!m)
T_{q\bar{q}}^{*}\Big]
\frac{C_{_{F}}}{N}\phi_{_A}^{q,q}
(\k_{2\perp})
\nonumber\\
&&
+\int_{\k_\perp}
\!
{\rm tr}
\Big[(\slq\!+\!m)T_{q\bar{q}}(\slp\!-\!m)
T_{g}^{*}\Big]
\phi_{_A}^{q\bar{q},g}
(\k_{2\perp};\k_\perp)+{\rm h.c.}
\nonumber\\
&&\;
+{\rm tr}
\Big[(\slq\!+\!m)T_{g}(\slp\!-\!m)T_{g}^{*}\Big]
\phi_{_A}^{g,g}(\k_{2\perp})
\Big\}
\varphi_p(\k_{1\perp})\; ,
\label{eq:cross-section-q}
\end{eqnarray}
A new distribution appears in the single quark cross-sec\-tion, defined
as:
\begin{equation}
\phi_{_A}^{q,q}(l_\perp)\equiv \frac{2\pi^2 R_{_A}^2 l_\perp^2}{g^2 N}
\int_{\x_\perp}
e^{i\l_\perp\cdot\x_\perp}
{\rm tr}\left<
{\wt U}(0)
{\wt U}^\dagger(\x_\perp)
\right>\; .
\end{equation}
Once again, strict $k_\perp$-factorization is broken: one needs three
different correlation functions in order to write this cross-section.

\section{Correlators of Wilson lines}
\label{sec:avg}
It is in general extremely difficult to calculate the correlators of
Wilson lines that appear in the previous formulae for the pair and
single quark cross-sections. In principle, one would have to solve the
JIMWLK equation, which is a question that so far has not received a
full (numerical or otherwise) answer, despite some encouraging
attempts \cite{RummuW1}.

One can however calculate these correlators analytically in the
McLerran-Venugopalan model where the distribution of color
sources in the nucleus has the Gaussian form
\begin{equation}
W[\rho]=\exp\left[
-\int_{\x_\perp}\frac{\rho_a(\x_\perp)\rho_a(\x_\perp)}{2\mu_{_A}^2}
\right]\; .
\end{equation}
Since this model is quite relevant at moderate values of $x$ for large
nuclei, the closed expressions one obtains from it are of interest in
assessing issues such as the magnitude of the breaking of
$k_\perp$-factorization, or in order to make phenomenological
predictions in this kinematical domain. The expressions of the various
correlators have been derived in \cite{BlaizGV2}. Let us denote
\begin{equation}
\Gamma(\x_\perp-\y_\perp)
\equiv g^4\int_{\z_\perp}\left[
G_0(\x_\perp-\z_\perp)-G_0(\y_\perp-\z_\perp)
\right]^2\; , 
\end{equation}
where $G_0$ is the free 2-dimensional propagator
\begin{equation}
G_0(\x_\perp-\z_\perp)\equiv
\int\frac{d^2\k_\perp}{(2\pi)^2} \;\;\frac{e^{i\k_\perp\cdot(\x_\perp-\z_\perp)}}
{\k_\perp^2}\; . 
\end{equation}
The simplest ones are the 2-point correlators
\begin{eqnarray}
&&
\left<U(0)U^\dagger(\x_\perp)\right>_{aa}
=(N^2-1)e^{-\frac{N}{2}\mu_{_A}^2 \Gamma(\x_\perp)}\; ,
\nonumber\\
&&
{\rm tr}\left<{\wt U}(0){\wt U}^\dagger(\x_\perp)\right>
= N e^{-\frac{C_f}{2}\mu_{_A}^2\Gamma(\x_\perp)}\; .
\end{eqnarray}
Since the 3- and 4-point correlators have quite complicated
expressions, we will only quote here the simplified results for the
large $N$ limit\footnote{It is possible, albeit quite cumbersome, to
  evaluate the 3-point function numerically without doing the large
  $N$ approximation \cite{FujiiGV1}.}
\begin{eqnarray}
&&
{\rm tr}\left<
{\wt U}(\x_\perp)t^a {\wt U}^\dagger(\y_\perp) t^b U_{ba}(0)
\right>
=
\nonumber\\
&&\qquad\qquad=
NC_f
e^{-\frac{N}{4}\mu_{_A}^2 [\Gamma(\x_\perp)+\Gamma(\y_\perp)]}\; ,
\nonumber\\
&&
{\rm tr}\left<
{\wt U}(\x_\perp)t^a {\wt U}^\dagger(\y_\perp){\wt U}(\v_\perp)t^a
{\wt U}^\dagger(\u_\perp)
\right>
=\nonumber\\
&&\qquad\qquad=
NC_f e^{-\frac{N}{4}\mu_{_A}^2[\Gamma(\x_\perp-\u_\perp)+\Gamma(\y_\perp-\v_\perp)]}\; .
\end{eqnarray}
One sees that the 3- and 4-point correlators are  expressed entirely 
in terms of two-point correlators in the large $N$ limit. This 
simplifies the computation  enormously.

\section{Discussion}
\label{sec:discussion}
\subsection{Generalities on $k_\perp$-factorization}
The two terms in eq.~(\ref{eq:Mf-final-1}) have a simple
interpretation. The first term, containing two Wilson lines in the
fundamental representation, corresponds to a process in which the
quark-antiquark pair is produced before the collision with the
nucleus. The pair then goes through the nucleus and scatters off the
color charges present in the nucleus. The second term, which contains
only one Wilson line in the adjoint representation, corresponds to a
process in which the gluon coming from the proton goes through the
nucleus and produces the quark-antiquark pair only after the collision
with the nucleus.

Thanks to the identity
\begin{equation}
{\wt U}(\x_\perp)t^a {\wt U}^\dagger(\x_\perp)
=
t^b U_{ba}(\x_\perp)
\end{equation}
between Wilson lines in the fundamental and adjoint representations,
one sees readily that $k_\perp$-factorization is recovered in the limit
where the difference $\x_\perp-\y_\perp$ between the transverse
coordinates of the quark and the antiquark is neglected. In other
words, $k_\perp$-factorization should be a valid approximation in any
kinematical regime where the typical transverse size of the pair is
small. This is expected to be the case if the quark mass is large, or
if the transverse momenta in the final state are large.

There is another way to reach the same conclusion from
eq.~(\ref{eq:Mf-final-1}). If the dependence of the factor
$T_{q\bar{q}}(\k_{1\perp},\k_\perp)$ on $\k_\perp$ is neglected, then
the integration over $\k_\perp$ in the amplitude produces a
$\delta(\x_\perp-\y_\perp)$. This automatically ensures
$k_\perp$-factorization. Since the typical $\k_\perp$ is of order
$Q_s$, it is possible to neglect $\k_\perp$ in the factor
$T_{q\bar{q}}$ if there is some other momentum scale (either $m$ or
$\q_\perp$) much harder than $Q_s$.

\subsection{Quark production}
The simplest physical observable to look at is the cross-section for
single quark production, obtained by integrating out the phase-space
of the anti-quark. This partonic cross-section can  be converted
into the cross-section for $D$ or $B$ mesons production, by
convolution with the appropriate fragmentation function. Since quite a
lot of phenomenology has been done in the framework of
$k_\perp$-fac\-to\-ri\-za\-tion, it is particularly interesting to investigate
the importance of the breaking of this factorization, and its
dependence on parameters such as the quark mass, the quark transverse
momentum, and the saturation scale in the nucleus.

A numerical computation of eq.~(\ref{eq:cross-section-q}) is under
way, and detailed results will be reported elsewhere \cite{FujiiGV1}.
One can nevertheless point to the following qualitative trends:
\begin{itemize}
\item The magnitude of the breaking of $k_\perp$-factorization
  decreases as the quark mass increases. Indeed, since the terms that
  break $k_\perp$-factorization correspond to extra rescatterings, it
  is natural that massive quarks are less sensitive to these effects
  than light quarks.
\item The magnitude of the breaking of $k_\perp$-factorization is
  maximal for a transverse momentum $q_\perp\sim Q_s$ of the quark, where $Q_s$ is the saturation scale in the nucleus. 
  One recovers $k_\perp$-factorization when the quark transverse momentum becomes
  much larger than all the other scales.
\item If $Q_s$ remains smaller or comparable to the quark mass and
  transverse momentum, the corrections due to the breaking of
  $k_\perp$-factorization enhance the cross-section. This is
  interpreted as a threshold effect: having more rescatterings tend to
  push a few more $Q\overline{Q}$ pairs just above the kinematical
  production threshold.
\item If $Q_s$ is large compared to the mass of the quark, then the
  corrections due to the breaking of $k_\perp$-factorization tend to
  reduce the cross-section at small transverse momentum. Since
  the typical mo\-men\-tum trans\-fer in a scattering is of the order of
  $Q_s$, it is indeed more difficult to produce light quarks with a small
  momentum if they scatter more.
\end{itemize}
In addition, it will be interesting to investigate the effect of this
breaking of $k_\perp$-factorization on the Cronin effect and its
rapidity dependence, in order to see whether the Cronin effect for
heavy quarks follows the same pattern as for gluon production
\cite{BlaizGV1,IancuIT2,KharzKT1}.

\subsection{Quarkonium production}
Quarkonium production in high-energy heavy ion collisions is a 
 key observable for diagnosing Quark Gluon Plas\-ma
formation\cite{MatsuS1}.  The anomalous suppression of the $J/\psi$,
compared with the baseline of the nuclear suppression, was indeed
observed at CERN-SPS\cite{Alessa1}.  In the experimental analysis, the
normal dissociation of the $J/\psi$ state by the nucleons is assumed
to be independent.  At higher energies, however, the nuclei in their
center-of-mass frame are Lorentz-contracted to be narrower than the
characteristic size of the quarkonium. Thus one expects the coherence between
the interactions with the nucleons to be more important.  In the
rest frame of a nucleus, on the other hand, the internal motion of the
heavy quarks is frozen during the interactions.

A model of the quarkonium projected on a target nucleus at high energy
is considered in \cite{FujiiM1,Fujii1}, which is formulated as an eikonal
propagation of the bound heavy quarks in the random color background
fields in the nucleus.  The multiple interactions with the color gauge
fields result in the random walk of the quarkonium state in the momentum
and color spaces.  The survival probability of the bound state, the
overlap of the initial and final pair state, is shown to be damped like a 
power-law rather than exponentially in the effective target
thickness $L$. This phenomenon is dubbed ``super-penetration'' in the QED
case.

In order to understand the baseline of the normal nuclear suppression
in the high-energy heavy ion collisions, it would be crucial to
investigate the nuclear effect on the production
process of  quarkonium rather than the attenuation of the
formed resonance\cite{QiuVZ1,Fujii2}. 
In this context, the analytic expression for the
quark production amplitude in the proton-nucleus collision within the MV model
will provide a good starting point for the further study of 
quarkonium formation. The multiple scattering effect in the nucleus
target is appropriately included in the framework, and it will
suppress the probability for producing the quark-pair
with  small relative momentum. We speculate that the formation 
probability of the small-momentum pair will be suppressed in some
power of the target size, and not exponentially, reflecting the random walk
nature of the multiple scatterings.

\section{Conclusions}
The calculation of the production quark-antiquark pairs in the
framework of the Color Glass Condensate, in the case of proton-nucleus
collisions, shows unambiguously that $k_\perp$-factorization is
broken. This is to be contrasted to what was observed for gluon
production where this factorization is preserved, despite the presence
of important rescattering effects in the nucleus. In the case of quark
production, one needs more correlators in order to describe the
nucleus. In particular, the cross-section depends on correlators
of  three and four Wilson lines. This opens the possibility of
obtaining more detailed information about correlations among  color
charges, as a function of rapidity. Conversely, experiments can test in detail 
CGC predictions for these higher point correlators.

\section*{Acknowledgements}
The work of one of us (R.V.) is supported in part  by DOE Contract
No. DE-AC02-98CH10886 and by a research   
grant from the Alexander Von Humboldt Foundation. He would like to
further acknowledge the kind hospitality of the Institute for  
theoretical physics at Bielefeld University and the Institute for
theoretical physics-II, at Hamburg University. 
The work of H.F.\ is supported in part by the Grants-in-Aid for Scientific
Research of Monbu-kagaku-sho (13440067, 16740132).
\bibliographystyle{unsrt}

\end{document}